\documentclass[a4paper]{jpconf}
\usepackage{graphicx}
\begin{document}
\title{From gluon topology to chiral anomaly: \\ Emergent phenomena in quark-gluon plasma}

\author{Jinfeng Liao}

\address{Physics Department and Center for Exploration of Energy and Matter,
Indiana University, 2401 N Milo B. Sampson Lane, Bloomington, IN 47408, USA.}
\address{RIKEN BNL Research Center, Bldg. 510A, Brookhaven National Laboratory, Upton, NY 11973, USA.}

\ead{liaoji@indiana.edu}

\begin{abstract}
Heavy-ion collision experiments at RHIC and the LHC have found a new emergent phase of QCD, a strongly coupled quark-gluon plasma (sQGP) that is distinctively different from either the low temperature hadron phase or the very high temperature weakly coupled plasma phase. Highly nontrivial emergent phenomena occur in such sQGP and  two examples will be discussed in this contribution: the magnetic component of sQGP that stems from topologically nontrivial configurations in the gluon sector; and the anomalous chiral transport that arises as macroscopic manifestation of microscopic chiral anomaly in the quark sector. For both examples, their important roles in explaining pertinent heavy-ion data will be emphasized.  
\end{abstract}

\section{Introduction}

For the overwhelming part of the pursue of science in the past two thousand years or so since the early philosophers, people were trying to understand the structure of matter by {\it ``reduction''}, i.e. by figuring out the most fundamental ``building blocks'' of all matter. That approach successfully led to the celebrated ``Standard Model'' and is leading further toward ideas beyond it. The opposite route, that is, to understand the integration of these ``building blocks'' into the various emergent forms and properties of matter, is no less fundamental and often proves equally nontrivial, as emphasized only relatively recently. In terms of the famous notion by P. W. Anderson, {\it ``More Is Different''}~\cite{Anderson:1972pca}. A simple example to illustrate this notion, is perhaps the ``lego matter'': there are just a few different types of basic pieces with rather simple ways of coupling any two pieces together; given many thousands of lego pieces, one could imagine a simple/natural phase of lego matter being just a big random pile of them; there are however all kinds of ordered structures from a toy car to a splendid castle that could be constructed given enough of these pieces --- these constructions are ``emergent phases'' of lego matter. The whole of condensed matter physics, is essentially studying the wealth of {\it emergent phenomena}  based on the fundamental force of electromagnetism.    

The world of nuclear matter is of course governed by a different type of fundament force, described by the theory of quantum chromodynamics (QCD).  Nevertheless as recently emphasized by F. Wilcezk:  ``The study of the strong interactions is now a mature subject --- we have a theory of the fundamentals (QCD)......The important questions involve `emergent phenomena'.'' Indeed most of today's nuclear physics, in a way, may be regarded as the {\it ``condensed matter physics of QCD''}, with several very active frontiers locating at different regimes on the QCD phase diagram in terms of temperature and baryon density. At very cold and dilute setting, the QCD matter is vacuum-like and one is at the {\it ``force frontier''}, studying the emergent nonperturbative force  in the vacuum-like environment that binds partons into observed hadrons by investigating spectroscopy, hadron scatterings/reactions, parton distributions, etc.  Moving into the cold but extremely baryon-rich regime, the QCD matter takes the form from large nuclei to dense nuclear matter and eventually toward dense quark matter. This is the {\it ``dense frontier''} where one investigates the many emergent phases that are crucial for understanding some of the most remarkable cosmic and astrophysical phenomena such as nucleosynthesis, supernovae and neutron stars.  Finally upon increase of  temperature to the extreme, the QCD matter enters an entirely new (and yet primordially old) phase as a quark-gluon plasma and one is at the {\it ``hot frontier''}, creating the cosmic epoch QGP and measuring its properties via heavy-ion collisions.   

For the rest of this contribution I will focus on the ``hot frontier'' and discuss a new emergent phase of QCD, the strongly coupled quark-gluon plasma (sQGP) that has been found in heavy-ion experiments at the Relativistic Heavy Ion Collider (RHIC) and the Large Hadron Collider (LHC)~\cite{Gyulassy:2004zy,Shuryak:2004cy}. This sQGP, spanning a temperature of range of about $1\sim 3$ times the parton/hadron transition temperature $T_c\sim 165\rm MeV$,  is markedly different from either the low temperature hadron phase or the very high temperature weakly coupled plasma phase. In particular I will provide  two nontrivial examples of emergent phenomena in sQGP that became appreciated only relatively lately.  The first example involves the magnetic component of sQGP that stems from topologically nontrivial configurations known as chromo-magnetic monopoles in the gluon sector, while the second example is about the anomalous chiral transport that arises as macroscopic manifestation of microscopic chiral anomaly in the quark sector. In both cases, the   importance in explaining pertinent heavy-ion data will be emphasized.

\section{Magnetic component of sQGP}
  
It was known long ago that the vacuum or vacuum-like state at $T\ll \Lambda_{QCD}$ is a complicated phase of QCD matter, with the remarkable phenomena of color confinement and spontaneous chiral symmetry breaking. It was also realized shortly after the birth of QCD that there should be {\it a natural/simple phase of QCD matter at asymptotically high temperature} $T\gg \Lambda_{QCD}$ where it should be much like a plasma of weakly coupled color charges (quarks and gluons). Indeed such early ideas motivated and eventually led to the successful heavy-ion experimental programs in the past decades. So where are we now?  In short, the QGP is created, but turns out to be a rather different one than the originally expected asymptotically free QGP. Evidences from heavy-ion data as well as from lattice QCD simulations, particularly from its nearly perfect fluidity and extreme color-opaqueness, have been accumulated in support of this finding. In hindsight, a strongly coupled QGP may not be that surprising: after all, the reachable temperature in current heavy-ion collisions is only around $T\sim$ few times $\Lambda_{QCD}$. The measured properties of the sQGP, such as 
the shear viscosity (normalized by entropy density) $\frac{\eta}{s}$ and the (normalized) jet transport coefficient $\frac{\hat{q}}{T^3}$, clearly indicate nonperturbative dynamics in this temperature regime and calls for a deeper theoretical understanding. In addressing such problem, it was realized and proposed that this quark-gluon plasma is a special form of emergent plasma which is in a post-confinement regime {\em above but close to   the parton/hadron phase boundary}, preserving essential ingredients that drive the system toward  the confinement transition at $T_c$~\cite{Liao:2006ry,Liao:2008jg,Liao:2012tw}. 

An obvious question is: what are the microscopic degrees of freedom  in this near-$T_c$ plasma? One hint comes from a comparison between how rapidly the thermal degrees of freedom (as measured by the normalized entropy density $s/T^3$) become liberated with increasing temperature and how rapidly the color degrees of freedom (as measured by Polyakov loop $<L>$, being zero in the confined phase while being unity in the fully deconfined phase) become liberated. Lattice data tell us that while the normalized entropy density increases by almost an order of magnitude in a very narrow window just around $T_c$, the Polyakov loop stays well below unity even at $2T_c$. This implies that significant suppression on (chromo-)electric charges (i.e. the quarks and gluons) persists even into the $1\sim 2\rm T_c$, as first emphasized with the name semi-QGP by Pisarski et al~\cite{Pisarski:2009zza}, and such non-perturbative suppression must originate from the same dynamics that enforces confinement below $T_c$. The ``mismatch'' between the liberation of thermal particles and the color charges in the plasma, must also indicate active alternative degrees of freedom that are not electrically charged and at play in the plasma. It was first conjectured by Shuryak and myself that a thermal {\it magnetic component} of relatively light and abundant (chromo-)magnetic monopoles must prevail in the near-$T_c$ plasma~\cite{Liao:2006ry,Liao:2008jg}. 
Indeed, if the confining regime below $T_c$ would be a (dual) superconducting phase of magnetic monopoles, then it would be natural to expect a {\em normal phase} of the same thermal ensemble of monopoles above $T_c$. These monopoles are themselves magnetically charged and free from the Polyakov loop suppression. They should be the origin of  the suppression of quarks/gluons in the semi-QGP regime and their Bose-Einstein condensation at $T_c$ signals the confinement transition. This is  the ``magnetic scenario'' for the sQGP. Evidences for such a picture were seen from lattice simulations for pure gauge  theories where these monopoles are identified (albeit in certain gauge choice) and counted, with their densities and mutual-correlations being measured at temperatures around and above $T_c$~\cite{Chernodub:2006gu,D'Alessandro:2007su,Bonati:2013bga}. 

It is natural to ask how the magnetic scenario may help understand the observed properties of the sQGP such as $\eta/s$ and $\hat{q}/T^3$ and whether the magnetic scenario is indispensable to explaining certain heavy-ion data. As pointed out in~\cite{Liao:2006ry,Liao:2008jg},  by virtue of the Dirac condition the scatterings between electric and magnetic components are always strong and frequent, which is precisely what an abundant magnetic component has to offer for making a nearly perfect fluid with extreme color opaqueness. Based on such reasoning, a nontrivial prediction for the jet quenching phenomenon was first made in \cite{Liao:2008dk}, namely the jet-medium interaction is strongly enhanced in the near-$T_c$ regime due to the magnetic component. The  analysis of the measured azimuthal anisotropy data for jet energy loss was shown to provide strong  phenomenological evidence for such a near-$T_c$ enhancement~\cite{Liao:2008dk,Zhang:2012ie} which implies a prominent peak around $T_c$ for the normalized jet transport coefficients $\hat{q}/T^3$ in analogy to the ``critical opalescence''.

\begin{figure}[!hbt]
\includegraphics[scale=0.45]{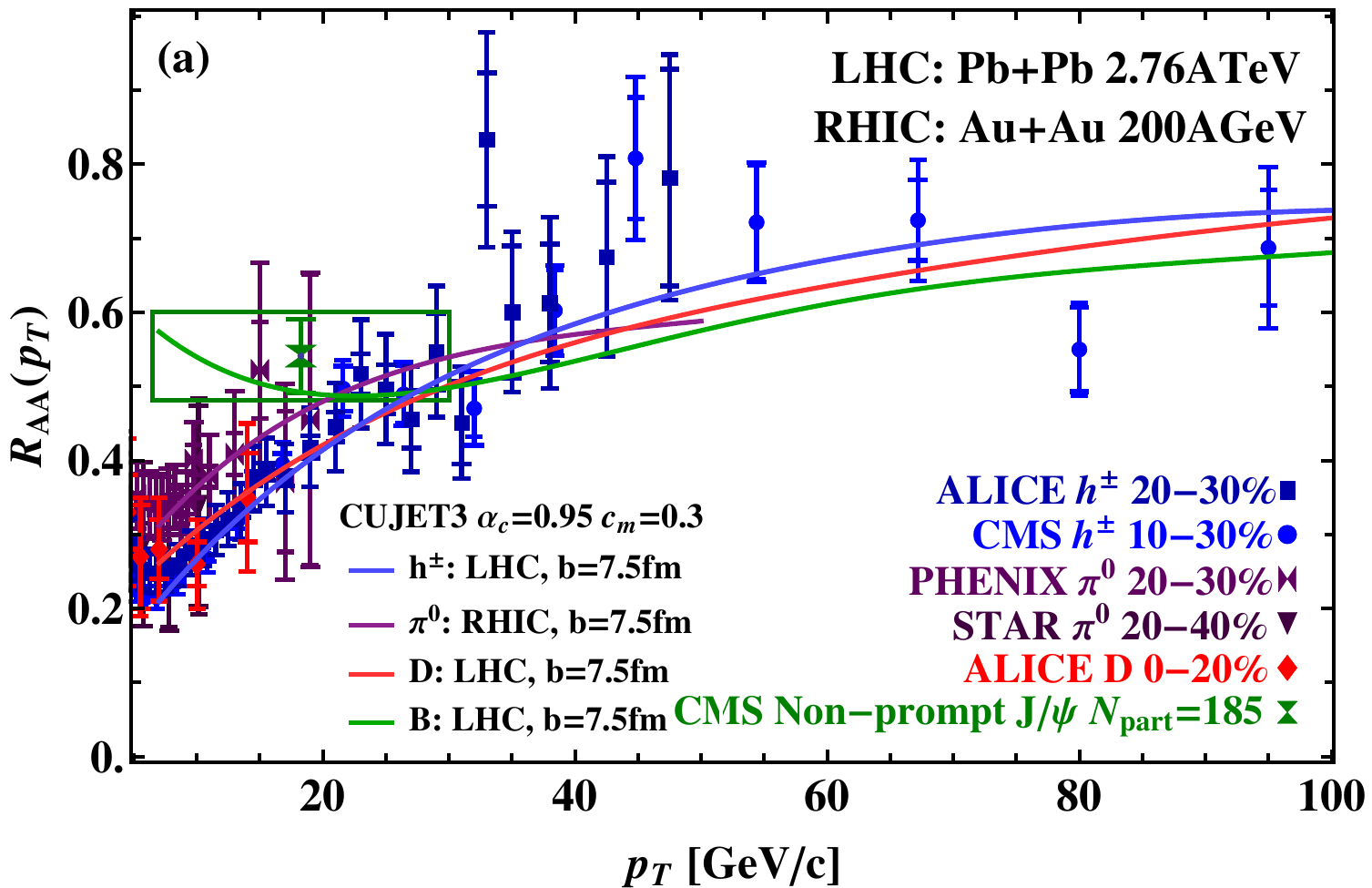} \hspace{2pc}%
\includegraphics[scale=0.45]{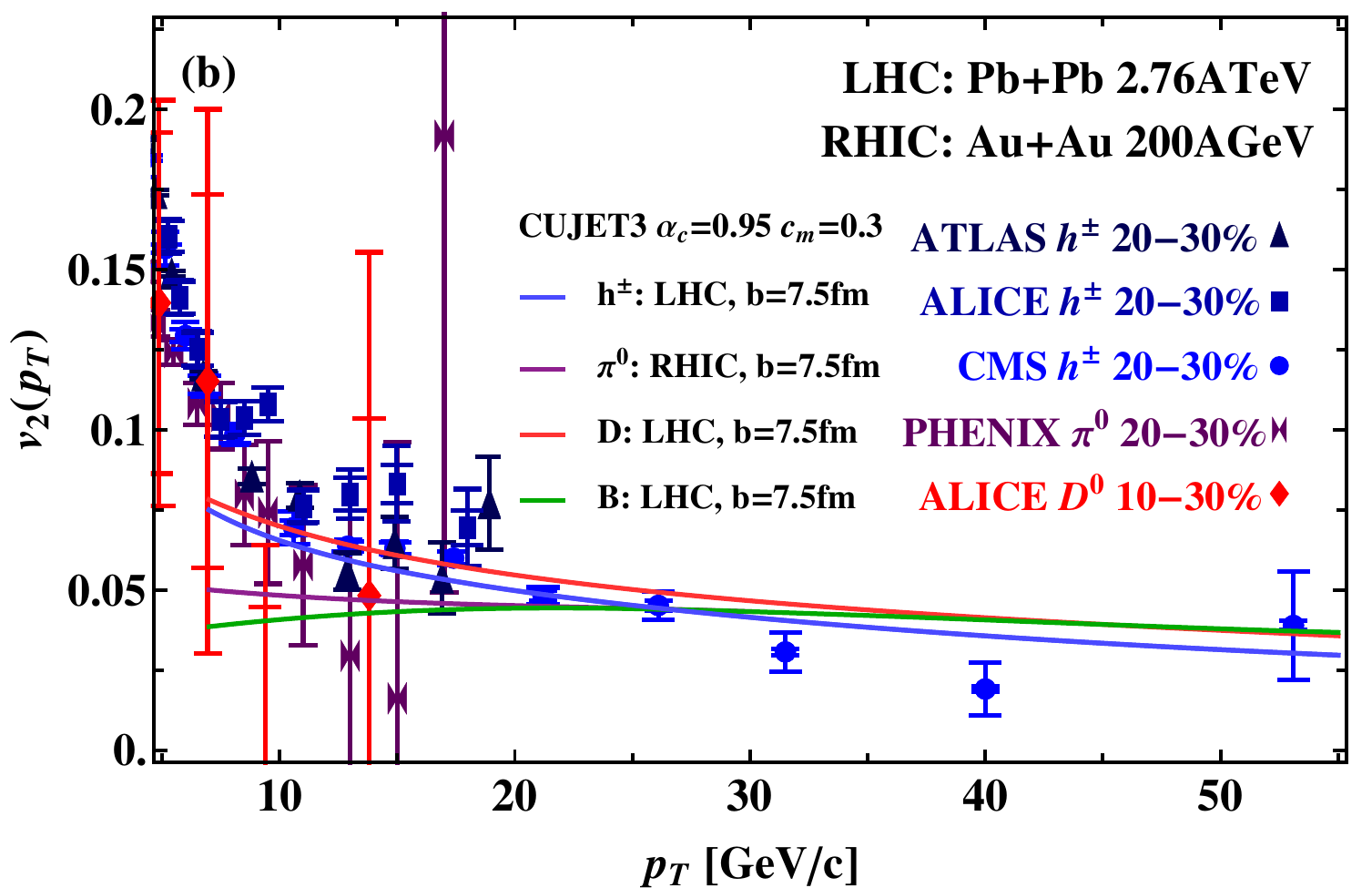}
\caption{\label{fig_CUJET}
(a) $R_{AA}(p_T)$ and (b) $v_{2}(p_T)$ for inclusive neutral pions ($\pi^0$) and charged particles ($h^\pm$) in Au+Au $\sqrt{s_{NN}}=200$ GeV and Pb+Pb $\sqrt{s_{NN}}=2.76$ TeV collisions as well as  that for open heavy flavors (D meson, red; B meson, green) at LHC semi-peripheral Pb+Pb $\sqrt{s_{NN}}=2.76$ TeV, computed from CUJET3.0 with the impact parameter $b=7.5$ fm, and compared with various available data from ALICE, ATLAS, CMS, PHENIX and STAR. See \cite{Xu:2014tda,Xu:2015bbz} for details.}  
\end{figure}

\begin{figure}[!hbt]
\includegraphics[scale=0.45]{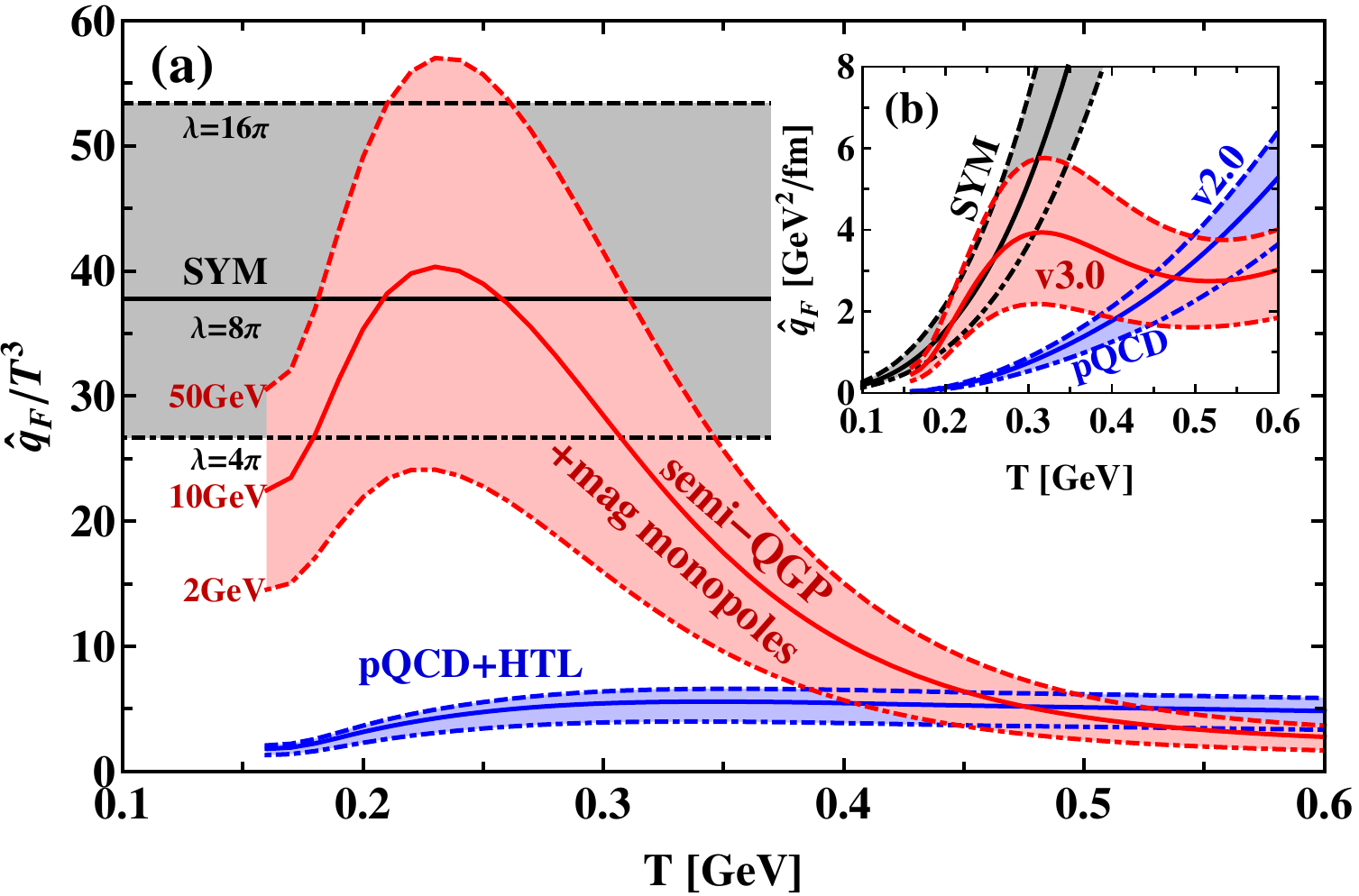} \hspace{2pc}%
\includegraphics[scale=0.45]{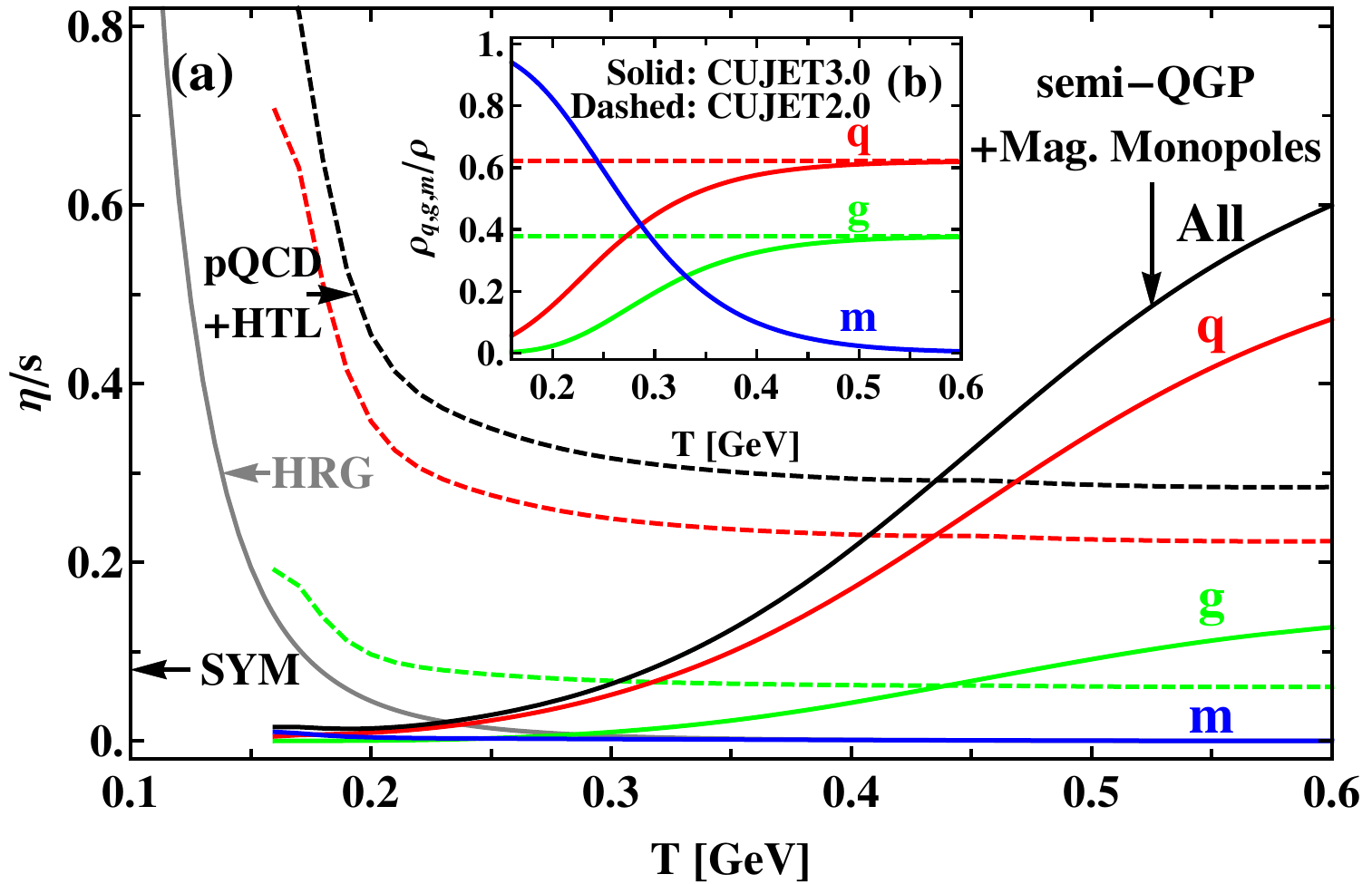}
\caption{\label{fig_eta_qhat}
Temperature dependence of  jet  transport coefficient $\hat{q}/T^3$  (left) and shear viscosity per entropy density $\eta/s$ (right), computed from CUJET3.0 (see \cite{Xu:2014tda,Xu:2015bbz} for details). }
\end{figure}

More recently, a major step forward was achieved for implementing the magnetic component of sQGP into a sophisticated and realistic modeling framework for heavy-ion collisions~\cite{Xu:2014tda,Xu:2015bbz}.  This  new comprehensive simulation framework, called CUJET3.0,  (1) treats the radiative energy loss in the DGLV formalism; (2) convolutes energy loss with a bulk evolution of heavy-ion collisions from the state-of-the-art viscous hydrodynamic simulation (VISHNU hydro) that is data-validated; (3) implements nonpeturbative near-Tc enhancement of jet-medium coupling by introducing the monopole component in the plasma constituents as well as the semi-QGP suppression on quarks and gluons; (4) constrains the thermodynamic contents of the plasma constituents by current lattice QCD data. Using the CUJET3.0,   jet quenching (leading-hadron) observables have been systematically investigated, including  the nuclear modification factor $R_{AA}$ and its azimuthal anisotropy $v_2$ at high transverse momenta for light and heavy flavors as well as for both RHIC and the LHC. Remarkably, this new framework with only one essential parameter successfully describes experimental data for  seven sets of single hadron observables, including light hadrons' $R_{AA}$ and $v_2$ at RHIC and LHC, D meson $R_{AA}$ and $v_2$ at LHC, as well as B meson $R_{AA}$ at LHC: see comparison in Fig.~\ref{fig_CUJET}.  This framework has also been used to predict  the temperature dependence of sQGP transport properties $\hat{q}/T^3$ and $\eta/s$. The results show a strong  non-monotonic structure, with a prominent peak for $\hat{q}/T^3$ and a strong minimum for $\eta/s$, both located   in the near-Tc regime: see Fig.~\ref{fig_eta_qhat}. The values of these coefficients from CUJET3.0 are in consistency with the phenomenological values inferred  from experimental data.

\section{Anomalous chiral transport}

Let me now move to the discussion of an emergent phenomenon in the quark sector. In the hot QGP the chiral symmetry is resorted and the light flavor quarks (u and d), with $m_{u,d}\ll T$, can be considered chiral fermions to very good approximation.  A most fundamental aspect in the quantum theory of chiral fermions is the famous chiral anomaly. An interesting question, is how such microscopic anomaly might manifest in macroscopic many-body setting. The answer by now is well known: from  chiral anomaly emerges the anomalous chiral transport processes, with the most notable example of chiral magnetic effect (CME). (See reviews in e.g. \cite{Kharzeev:2015znc,Liao:2014ava}.)  

As the QGP evolution in heavy-ion collisions is well described by hydrodynamics, let me rephrase this problem in the hydrodynamic context. The hydrodynamics by itself is precisely an emergent phenomenon from microscopic symmetries in the long-time and long-distance limit: spacetime translational invariance $\to$ conserved energy-momentum in microscopic scatterings $\to $ macroscopic hydro equations $\partial_\mu T^{\mu \nu}=0$;  phase invariance $\to$ conserved charges in microscopic scatterings $\to $ macroscopic hydro equations $\partial_\mu J^{\mu}=0$. One may then ask: what happens in hydro for a ``half'' symmetry i.e. an anomaly that is conserved classically but broken quantum mechanically? The answer is a new type of anomalous hydrodynamics~\cite{Son:2009tf}, where the chiral anomaly leads to new type of hydro equations for chiral currents: $\partial_\mu J^{\mu}_{R/L}= \pm C_A E^\mu B_\mu$ (where $C_A$ the universal anomaly coefficient), and more importantly a constituent relation $J^\mu = nu^\mu + \nu^\mu_{NS} + \xi \omega^\mu + \xi_B B^\mu$ with the second term the normal Navier-Stocks viscous term while the last two terms the anomalous chiral transport currents corresponding to the chiral vortical effect and chiral magnetic effect respectively. 
This remarkable hydrodynamic framework that distinguishes ``right'' and ``left'', provides the necessary tool for quantitatively modeling anomalous chiral transport effects in heavy-ion collisions.  

\begin{figure*}
\begin{center}
\includegraphics[scale=0.25]{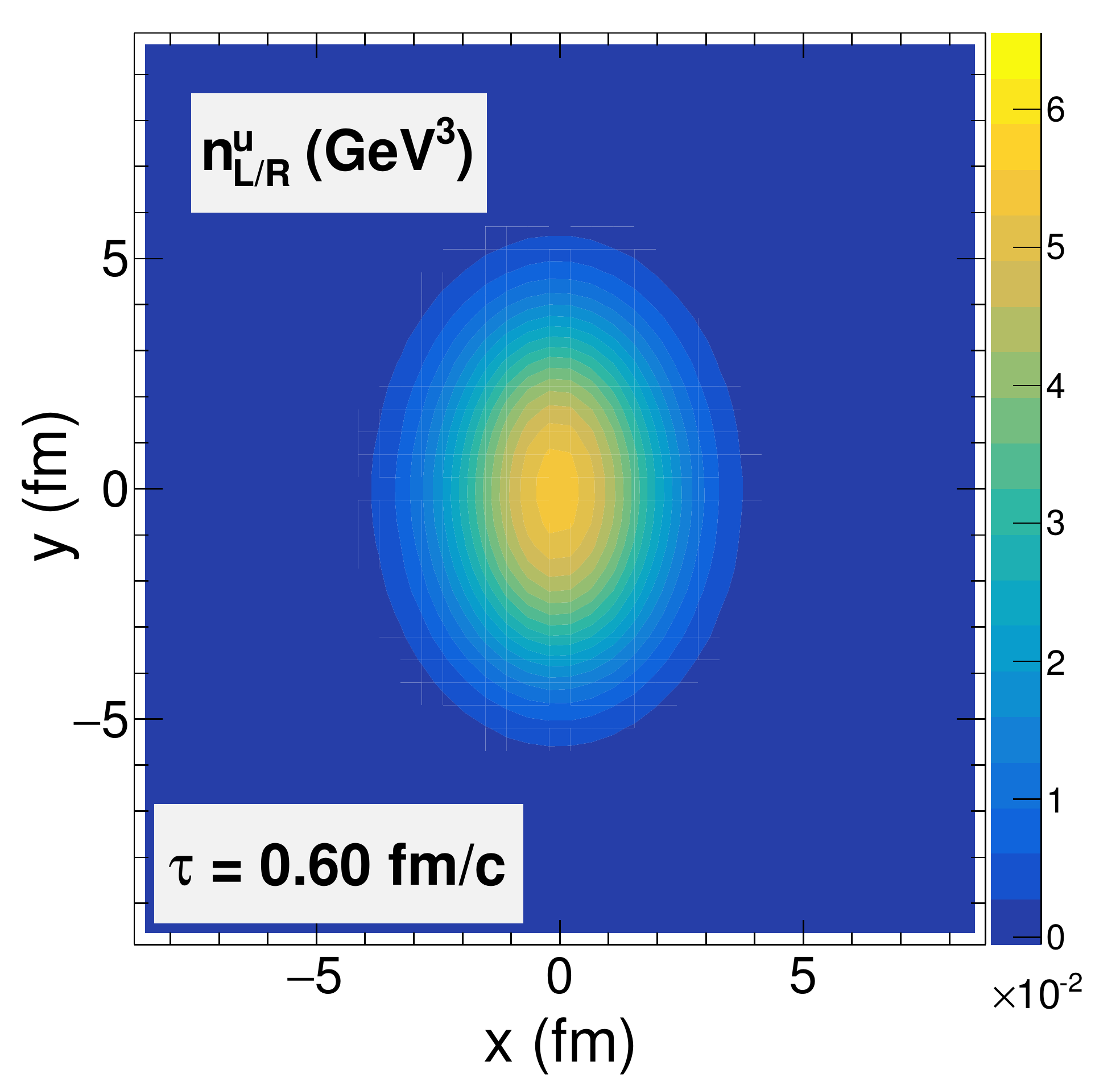} \hspace{2pc}
\includegraphics[scale=0.25]{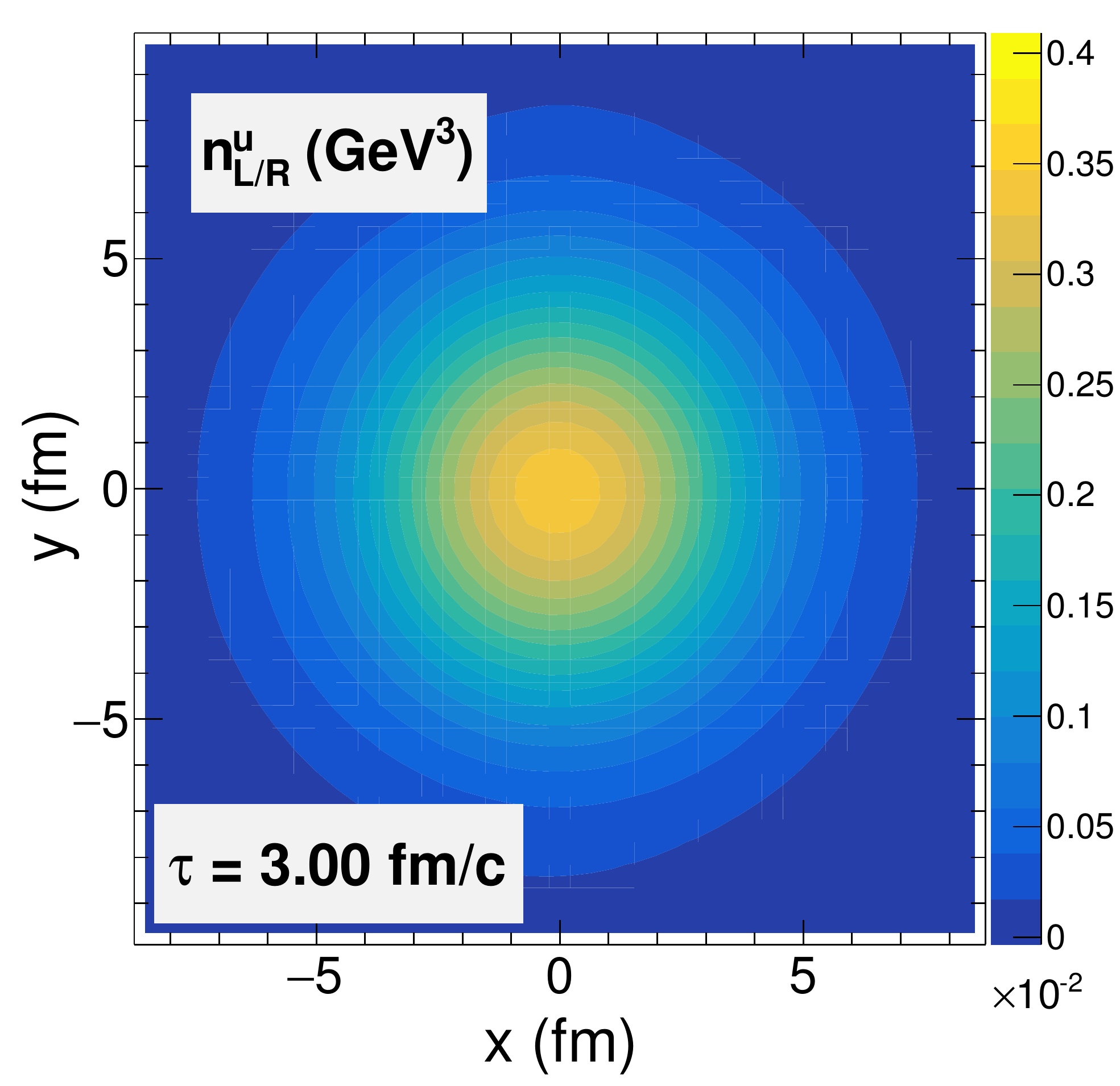} \\
\includegraphics[scale=0.25]{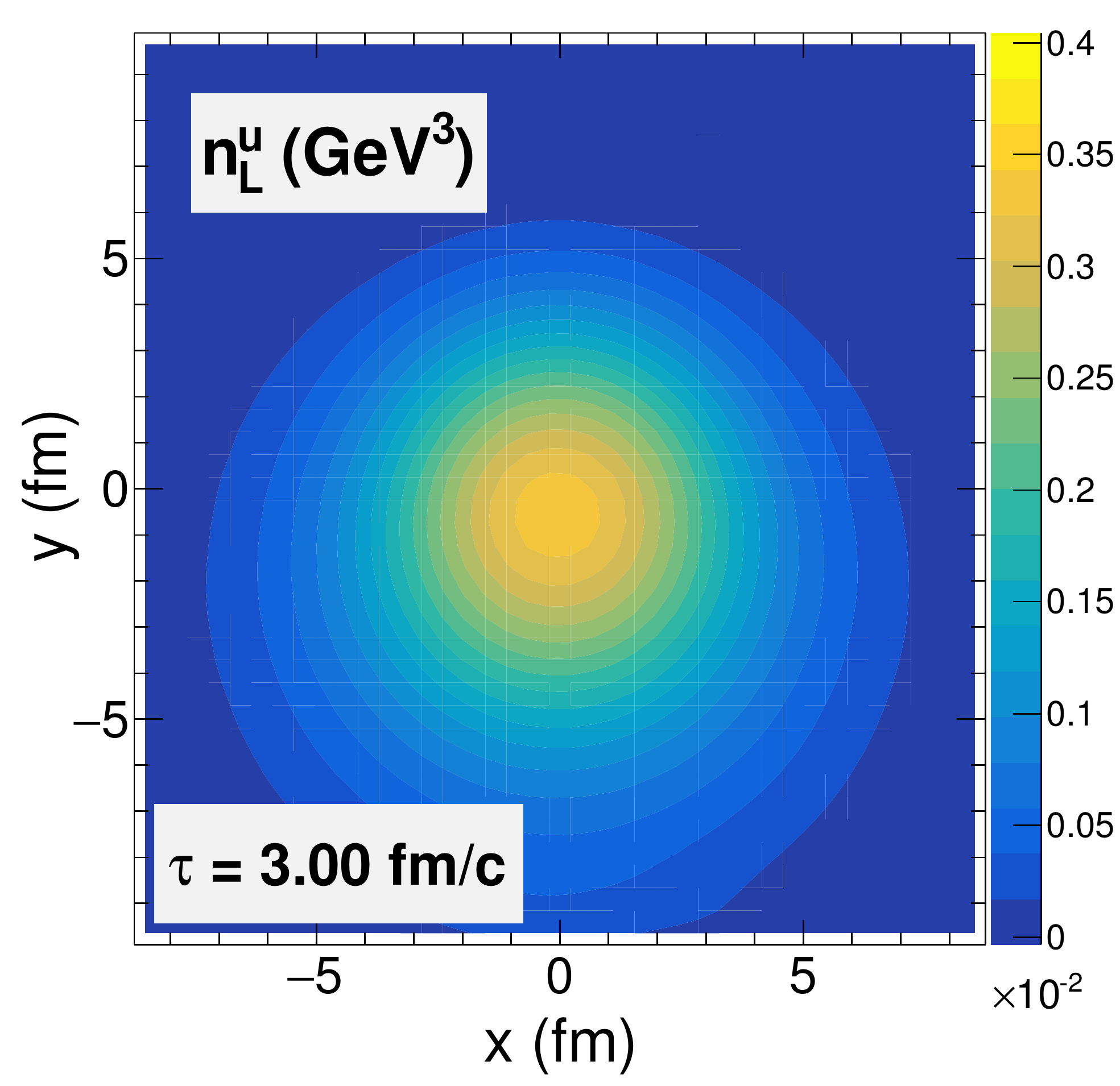} \hspace{2pc}
\includegraphics[scale=0.25]{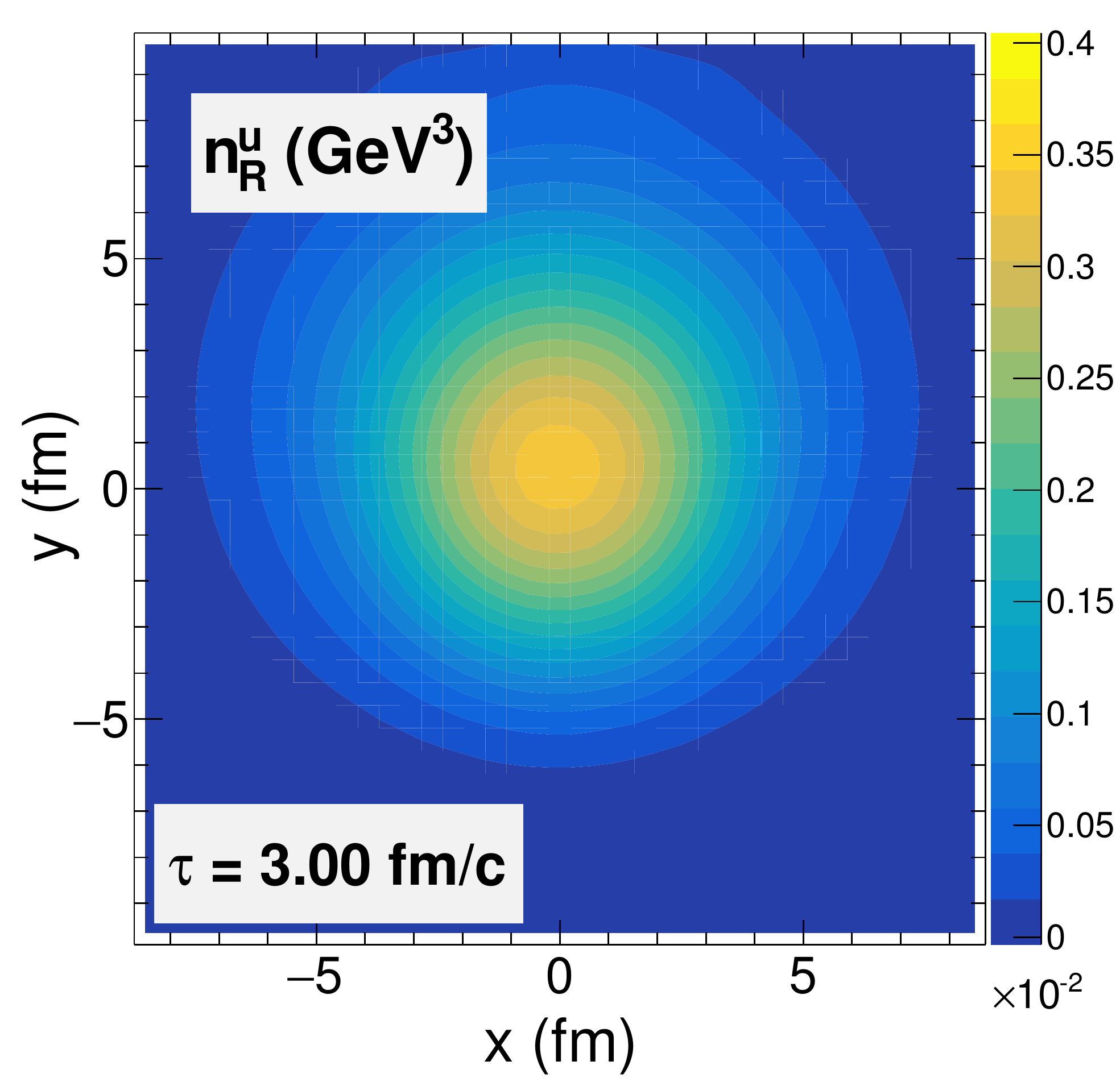}
\caption{Charge densities at 3~fm/$c$ starting from the same initial conditions (upper-left) in  Anomalous-Viscous Fluid Dynamics~\cite{Jiang:2016wve} for three cases: (upper-right) without $\vec B$ field; (lower-left) with $\vec B=B\hat{y}$ for left-handed density;  (lower-right) with $\vec B=B\hat{y}$ for right-handed density.  } \vspace{-0.3in}
\label{fig_density}
\end{center}
\end{figure*}

\begin{figure}
\includegraphics[width=3.2in,height=1.8in]{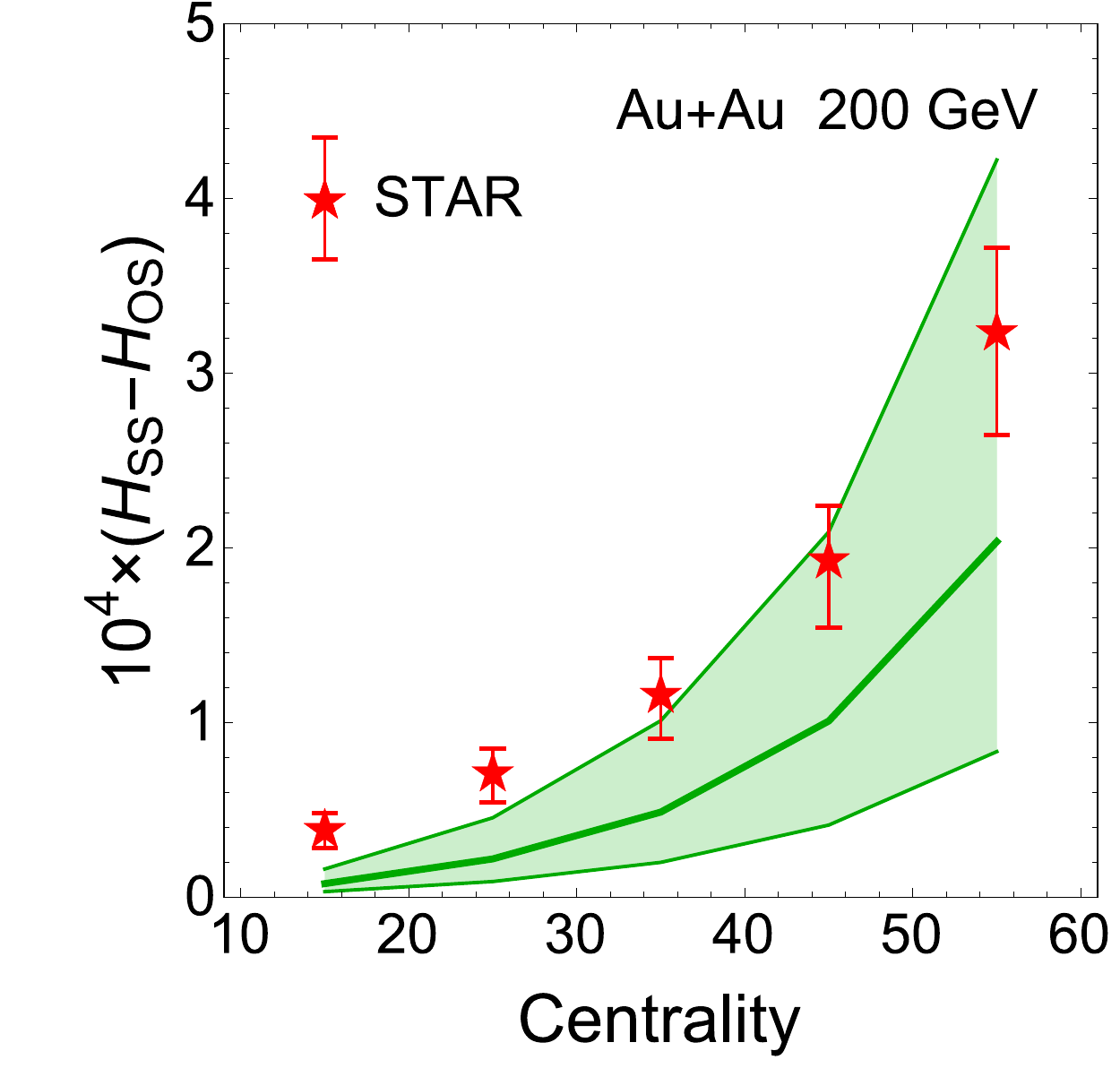} \hspace{2pc}%
\begin{minipage}[b]{14pc}
\begin{center}\caption{\label{fig_H} Quantitative predictions from Anomalous-Viscous Fluid Dynamics simulations for the CME-induced H-correlator~\cite{Jiang:2016wve}, in comparison with STAR measurements~\cite{Adamczyk:2014mzf}. The green bands reflect current theoretical uncertainty in the initial axial charge generated by gluonic field fluctuations.}
\end{center}
\end{minipage}
\end{figure}

In heavy-ion collisions large electric and magnetic fields are present due to fast moving ions. Here let me refrain from discussing electric field (which also induces interesting effects e.g. \cite{Huang:2013iia}) but focus on magnetic field $\vec B$, which has been quantified in its magnitude as well as azimuthal orientation with event-by-event simulations~\cite{Bloczynski:2012en}. The $\vec B$-driven CME could induce a charge separation effect along the out-of-plane direction, which can be measured with suitable charge-dependent azimuthal correlations. Such measurement faces the challenge of significant background contributions but a recent STAR analysis~\cite{Adamczyk:2014mzf} has separated the potential signal via a two-component subtraction method~\cite{Bzdak:2012ia}.  Quantitative predictions for CME signal, based on realistic bulk evolution and integrating anomalous hydro framework, become crucial for comparison with data (see early attempts in e.g.~\cite{Hirono:2014oda,Yin:2015fca}). A tool for this purpose is now being built, the Anomalous-Viscous Fluid Dynamics (AVFD) simulation~\cite{Jiang:2016wve} which solves the anomalous hydro equations with 2nd-order viscous terms for chiral fermion currents in QGP with magnetic fields, on top of the data-validated VISHNew~\cite{Shen:2014vra} bulk hydro evolution. The essential features of chiral density evolution in AVFD can be demonstrated in Fig.~\ref{fig_density} where anomalous chiral transport is evident and differentiates the R/L fermions. Finally the quantitative predictions for CME-induced correlations from AVFD simulations, with realistic magnetic field magnitude and lifetime as well as estimated initial axial charge density,  are presented in Fig.~\ref{fig_H} and compared with pertinent STAR data. The main conclusion is that up to theoretical uncertainty, the CME predictions are quantitatively in the right ballpark of observed correlations. The residue difference might originate from a number of sources: theoretically various fluctuations  and resonance decays need to be further included, while experimentally there are further backgrounds that shall be suitably suppressed in the H-correlators. These expected improvements together with anticipated isobaric collisions shall allow an ultimate conclusion.

 \ack{The author thanks M. Gyulassy, Y. Jiang, S. Shi, J. Xu, Y. Yin for collaborations on the results reported here. The author is grateful to H. Huang, R. Pisarski, E. Shuryak, G. Wang and N. Xu for discussions. This material is based upon work supported by the U.S. Department of Energy, Office of Science, Office of Nuclear Physics, within the framework of the Beam Energy Scan Theory (BEST) Topical Collaboration. The work is also supported in part by the National Science Foundation (Grant No. PHY-1352368) and by the RIKEN BNL Research Center. }

\end{document}